\documentclass[aps,pra,twocolumn,showpacs]{revtex4-1}

\usepackage{amsmath,amssymb}
\usepackage{graphicx}

\usepackage{color}

\graphicspath{ {.} {./pics/} }

\newcommand{\rmi}{\mathrm{i}}
\newcommand{\rmd}{\mathrm{d}}
\newcommand{\rme}{\mathrm{e}}

\begin{document}

\title{Strongly entangled light from planar microcavities}

\author{D. Pagel}
\affiliation{Institut f\"ur Physik, Ernst-Moritz-Arndt-Universit\"at Greifswald, 17487 Greifswald, Germany}
\author{J. Sperling}
\affiliation{Institut f\"ur Physik, Universit\"at Rostock, 18051 Rostock, Germany}
\author{H. Fehske}
\affiliation{Institut f\"ur Physik, Ernst-Moritz-Arndt-Universit\"at Greifswald, 17487 Greifswald, Germany}
\author{W. Vogel}
\affiliation{Institut f\"ur Physik, Universit\"at Rostock, 18051 Rostock, Germany}

\begin{abstract}
	The emission of entangled light from planar semiconductor microcavities is studied and the entanglement properties are analyzed and quantified.
	Phase-matching of the intra-cavity scattering dynamics for multiple pump beams or pulses, together with the coupling to external radiation, leads to the emission of a manifold of entangled photon pairs.
	A decomposition of the emitted photons into two parties leads to a strong entanglement of the resulting bipartite system.
	For the quantification of the entanglement, the Schmidt number of the system is determined by the construction of Schmidt number witnesses.
        It is analyzed to which extent the resources of the originally strongly entangled light field are diminished by dephasing in propagation channels.
\end{abstract}

\pacs{
	03.67.Bg,
	03.67.Mn, 
	42.50.Dv, 
	71.36.+c  
}

\maketitle

\section{Introduction}
	The interaction of light and matter is a fundamental issue connecting elements of quantum optics and solid state physics.
	It offers a wide range of quantum effects, for example, the emission of various kinds of nonclassical light.
	These phenomena sensitively depend on the interaction of light with the fundamental excitations of the crystal.
	The nonclassical correlations of such systems can be used for various applications, such as quantum information processing, quantum metrology, and quantum communication (see, e.\,g.,~\cite{HHHH09,NC10}).

	One of the most prominent quantum phenomena is entanglement.
	It has been studied since the very first ideas of non-local superpositions of wave functions arose~\cite{EPR35,Sch35}.
	Entanglement has been used to perform a number of classically impossible operations in theory and experiment, such as, quantum teleportation, secure communication, and distillation protocols~\cite{HHHH09,NC10}.
	The latter ones require copies of entangled mixed states to distill pure entangled states~\cite{BBPSSW96,HSDFS10}, namely Bell states~\cite{Bel64}.
	One problem is the feasibility of appropriate quantum memories to store and manipulate the individual, entangled copies~\cite{HSP10,JWKFNSOPWP11}.
	Despite this, the determination of entanglement of, in general, mixed quantum states is still a challenging task.

	Typically, quantum correlations are determined from measurements of correlation functions.
	Here, we aim to quantify the measured correlations in terms of entanglement.
	The tricky relation between entanglement and correlations was mainly analyzed for spin systems~\cite{GBF03,VPC04}. 
Various entanglement measures have been introduced~\cite{VPRK97,AFOV08} and
        compared numerically and analytically~\cite{EP99,MG04,PV07,SV10pre}.
	It has been shown that the evaluation of an entanglement measure, especially for mixed states beyond qubits, is a sophisticated problem.

	In the first instance, it is convenient to quantify entanglement for pure states only.
	One example is the Schmidt number (SN)~\cite{TH00,SBL01,SV11b}.
	For pure entangled states the SN counts the number of required superpositions of local product states to express the given state.
	A generalization to mixed quantum states can be achieved by a convex roof construction~\cite{Uhl98}.
	The SN of a general quantum state can be determined by making use of the method of SN witnesses~\cite{TH00,BCHHKLS02}.
	Recently, an approach based on generalized eigenvalue equations---so-called SN eigenvalue equations---led to a general construction scheme for SN witnesses~\cite{SV11a}.
	Note that such an approach does not exist for other entanglement measures.
	Another advantage of the determination of the SN via SN witnesses is that these witnesses represent experimentally accessible observables.

	Common approaches for the generation of bipartite entangled states consider type-II parametric down conversion~\cite{KMWZ95} or biexciton decay in quantum dots~\cite{BSPY00,HSK03}.
	Another prominent example is based on parametric phenomena in two-dimensional semiconductor microcavities~\cite{WNIA92,HWSOPI94,Ciu04,Lang04,CBC05,SDSSL05,DTDCLBRD06,PDSSS09}.
	They are known to realize a strong coupling between cavity photons and excitons~\cite{HWSOPI94} resulting in an anticrossing of the mixed exciton-photon modes, called lower and upper polariton branches.
	The ground state of the polaritons has been studied with respect to general quantum properties~\cite{CBC05} and entanglement~\cite{AB12}.
	Stimulated scattering processes of polaritons within the lower branch have been shown to result in a large angle-resonant amplification of the pump field~\cite{SBSSWR00,CSDQ00} and to produce polarization entangled polariton pairs~\cite{SDSSL05,PDSSS09}.
	Scattering processes involving both polariton branches can lead to the emission of photon pairs, which are entangled with respect to the branch index~\cite{Ciu04,CBC05}.

	In the present work we show that semiconductor microcavities can be used to generate strongly entangled photons and demonstrate how their entanglement can be identified.
	In our study, we apply different pump beams to the microcavity, which leads to the emission of a large number of entangled photon pairs.
	These pair correlations can be identified as a strong entanglement, if we decompose the emitted light into two ensembles of beams.
	The identification of these strong correlations is done using SN witnesses.
	We quantify the impact of a lossy channel on the strongly entangled systems by determining the SN.
	This procedure is closely connected to the solution of the SN eigenvalue equations.
	As a result, we show that different degrees of dephasing require different kinds of witnesses to detect strong entanglement.

	We proceed as follows.
	In Sec.~\ref{Sec:PlanarMC}, we briefly recapitulate the physical description of the polariton formation in planar microcavities.
	The intra-cavity scattering dynamics leads to branch entangled polariton pairs, which will be discussed in Sec.~\ref{Sec:BEPP}.
	The coupling of the polaritons to radiation modes considered in Sec.~\ref{Sec:FreqEntPhot} yields strongly frequency-entangled photons.
	We verify their correlations by the use of entanglement and SN witnesses.
	In Sec.~\ref{Sec:LinDephase}, we study the dephasing due to the propagation of the frequency-entangled radiation through a linear dispersive medium.
	Section~\ref{Sec:SaC} presents our conclusions.

\section{Planar microcavity model}\label{Sec:PlanarMC}
	In this section, we briefly recapitulate the quantum Hamiltonian model for semiconductor microcavities.
	It is based on the bosonic picture of interacting excitons~\cite{TY99,CSQ01,Ciu04} and can easily be used to investigate polariton parametric scattering in momentum space.
	Another common approach is based on the dynamics-controlled truncation formalism~\cite{AS94,SG96,PDSSRG08b} that can be written in terms of the $T$ matrix~\cite{TKRKGB02}.

	In semiconductors the fundamental excitations are electron-hole pairs with radius $R_X$ and binding energy $E_b = e^2 / (2 \epsilon R_X)$, with $\epsilon$ being the static dielectric constant of the crystal.
	Since excitons are composite particles made up of fermions, they have an internal structure. 
	Moreover, we have to take into account an effective exciton-exciton interaction~\cite{TY99,CSQ01},
	\begin{equation}\label{HXX}
		H_{XX} = 6 E_b \frac{R_X^2}{A} \sum_{\mathbf{k},\mathbf{k'},\mathbf{q}} b_{\mathbf{k}+\mathbf{q}}^\dagger b_{\mathbf{k'}-\mathbf{q}}^\dagger b_\mathbf{k}^{} b_\mathbf{k'}^{} \:.
	\end{equation}
	In this equation $b_{\mathbf{k}}^{}$ ($b_{\mathbf{k}}^\dagger$) are bosonic annihilation (creation) operators of excitons with wave vector $\mathbf{k}$ and dispersion $E_X(\mathbf{k})$, and $A$ is the sample surface.
	Since we consider planar microcavities, all wave vectors in Eq.~\eqref{HXX} shall be in-plane.
	As a simplification, we assume dispersionless excitons $E_X(\mathbf{k}) = E_X$ and work in units where $\hbar = c = 1$.

	Coupling the excitons of the crystal to in-plane cavity photons with dispersion
	\begin{equation}
		E_C(k) = E_C(|\mathbf{k}|) = E_C(0) \sqrt{1 + (k/k_0)^2}\:,
	\end{equation}
	where $k_0 = E_C(0)$, we have to consider the exciton-photon interaction.
	The harmonic part of this interaction is given by
	\begin{equation}
		H_{XC} = \Omega_R \sum_\mathbf{k} b_\mathbf{k}^\dagger a_\mathbf{k}^{} + \text{H.\,c.} \:,
	\end{equation}
	where $2 \Omega_R$ denotes the vacuum Rabi-splitting and $a_\mathbf{k}^{}$ ($a_\mathbf{k}^{\dagger}$) are bosonic annihilation (creation) operators of the cavity photons with in-plane wave vector $\mathbf{k}$.
	It leads to lower ($j = 1$) and upper ($j = 2$) polariton branches
	\begin{equation}
		E_j(\mathbf{k}) = \frac{1}{2} \Big( E_C(k) + E_X \mp \sqrt{ \big( E_C(k) - E_X \big)^2 + 4 \Omega_R^2} \Big) \:,
	\end{equation}
	which depend on the modulus $k = |\mathbf{k}|$ only.

\begin{figure}
		\includegraphics[width=0.48\textwidth]{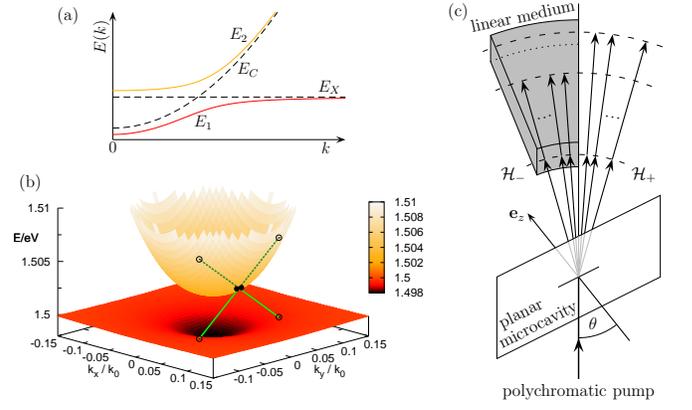}
		\caption{
			(Color online) Sketch of the considered physical processes.
			The inset (a) shows the dispersion relations of the excitons, cavity photons, and polaritons.
			Part (b) visualizes the interbranch polariton pair scattering.
			Panel (c) depicts the emission and propagation of the emitted entangled light.
		}
		\label{sketch}
	\end{figure}

	Figure~\ref{sketch}(a) schematically shows the polariton dispersions $E_1(\mathbf{k})$ and $E_2(\mathbf{k})$ (solid lines) as well as the dispersions $E_C(k)$ and $E_X$ of the cavity photons and the excitons (dashed lines).
	Note the anticrossing of the polariton branches, which is due to the strong coupling of exciton and cavity photon modes.
	The parameter $2 \Omega_R$ is oftentimes called polariton splitting, since it determines the distance $E_2(\mathbf{k}) - E_1(\mathbf{k})$ when the exciton and cavity photon modes are resonant, $E_C(k) = E_X$.

	The anharmonic part of the total Hamiltonian of the exciton-photon interaction (saturation) is~\cite{TY99,CSDQ00}
	\begin{equation}\label{HXC}
		H_{XC}^{\text{sat}} = - \frac{1}{2} \sum_{\mathbf{k},\mathbf{k'},\mathbf{q}} \frac{\Omega_R}{n_{\text{sat}} A} a_{\mathbf{k}+\mathbf{q}}^\dagger b_{\mathbf{k'}-\mathbf{q}}^\dagger b_\mathbf{k}^{} b_\mathbf{q}^{} + {\text{H.\,c.}} \;,
	\end{equation}
	where $n_{\text{sat}} = 7 / (16 \pi R_X^2)$ is the exciton saturation density.
	Together with the exciton-exciton interaction $H_{XX}$ it gives rise to an effective polariton-polariton interaction
	\begin{equation}\label{HPP}
		H_{PP} = \frac{1}{2} \sum_{\mathbf{k},\mathbf{k'},\mathbf{q}} \sum_{\substack{j_1,j_2,\\j_3,j_4}} \frac{R_X^2}{A} V_{\mathbf{k},\mathbf{k'},\mathbf{q}}^{j_1j_2j_3j_4} p_{j_1 \mathbf{k}+\mathbf{q}}^\dagger p_{j_2 \mathbf{k'}-\mathbf{q}}^\dagger p_{j_3 \mathbf{k}}^{} p_{j_4 \mathbf{k'}}^{} \:.
	\end{equation}
	Here the $p_{j \mathbf{k}}^{}$ ($p_{j \mathbf{k}}^\dagger$) are bosonic annihilation (creation) operators of polaritons in the lower or upper branch with in-plane wave vector $\mathbf{k}$.
	The effective branch-dependent potential $V_{\mathbf{k},\mathbf{k'},\mathbf{q}}^{j_1j_2j_3j_4}$ can be calculated through a unitary Hopfield transformation~\cite{Hop58}
	\begin{equation}\label{Hop}
		\begin{pmatrix} b_\mathbf{k} \\ a_\mathbf{k} \end{pmatrix} = \begin{pmatrix} M_{11\mathbf{k}} & M_{12\mathbf{k}} \\ M_{21\mathbf{k}} & M_{22\mathbf{k}} \end{pmatrix} \begin{pmatrix} p_{1\mathbf{k}} \\ p_{2\mathbf{k}} \end{pmatrix}
	\end{equation}
	as
	\begin{multline}\label{eff_potV}
		\frac{V_{\mathbf{k},\mathbf{k'},\mathbf{q}}^{j_1j_2j_3j_4}}{E_b} = 12 M_{1j_1 \mathbf{k}+\mathbf{q}} M_{1j_2 \mathbf{k'}-\mathbf{q}} M_{1j_3 \mathbf{k}} M_{1j_4 \mathbf{k'}} \\
		- \frac{8 \pi}{7} p_s \Big( M_{2j_1 \mathbf{k}+\mathbf{q}} M_{1j_2 \mathbf{k'}-\mathbf{q}} M_{1j_3 \mathbf{k}} M_{1j_4 \mathbf{k'}} \\
		+ M_{2j_4 \mathbf{k'}} M_{1j_3 \mathbf{k}} M_{1j_2 \mathbf{k'}-\mathbf{q}} M_{1j_1 \mathbf{k}+\mathbf{q}} \Big) \:.
	\end{multline}
	In Eq.~\eqref{eff_potV} we have introduced the ratio of polariton splitting to binding energy, $p_s = 2 \Omega_R / E_b$.
	For the matrix elements of the Hopfield transformation one finds the relations 
\begin{eqnarray}
M_{22\mathbf{k}} &=& M_{11\mathbf{k}} = 1 / \sqrt{1 + \rho_\mathbf{k}^2}\,,\\ 
M_{12\mathbf{k}} &=& - M_{21\mathbf{k}} = \sqrt{1 - M_{11\mathbf{k}}^2}\,,
\end{eqnarray} 
where
	\begin{equation}
		\rho_\mathbf{k} = \frac{E_2(\mathbf{k}) - E_C(k)}{\Omega_R} \:.
	\end{equation}
	Note that in contrast to the relations used in Ref.~\cite{Ciu04} the coefficient $M_{12\mathbf{k}}$ is always positive.

\begin{figure}
		\includegraphics[width=0.48\textwidth]{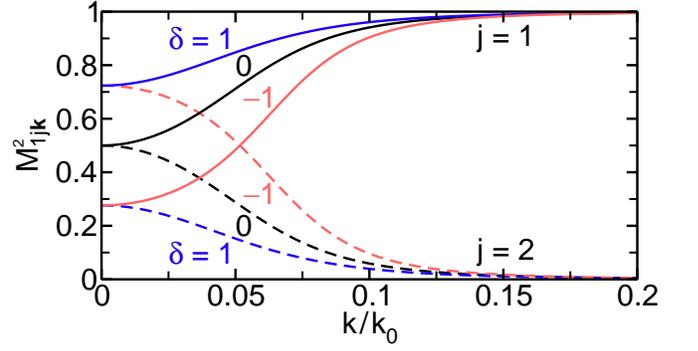}
		\caption{
			(Color online) Coefficients of the Hopfield transformation matrix in Eq.~\eqref{Hop}, for the parameters $E_C(0) = 1.5\,$eV and $\Omega_R = 2\,$meV.
			The solid lines correspond to $M_{11\mathbf{k}}^2$ and the dashed ones to $M_{12\mathbf{k}}^2$.
			The color of the curves indicates the values of the normalized detuning $\delta$.
			For large values of $|\mathbf{k}|$ the coefficient $M_{11\mathbf{k}}^2$ converges to one and $M_{12\mathbf{k}}^2$ vanishes.
		}
		\label{hopcoef}
	\end{figure}

	In Fig.~\ref{hopcoef} we show the dependence of the squared coefficients $M_{11\mathbf{k}}^2$ and $M_{12\mathbf{k}}^2$ on the modulus $k$ of the wave vector $\mathbf{k}$ for different values of the normalized detuning
	\begin{equation}
		\delta = \frac{E_C(0) - E_X}{2 \Omega_R} \:.
	\end{equation}
	For large values of $k$ the coefficient $M_{11\mathbf{k}}^2 \to 1$, and consequently excitons and cavity photons do not mix.
	The polariton modes are equal to the separated exciton and cavity photon modes.
	For smaller $k$ the value of $M_{11\mathbf{k}}^2$ depends on the detuning $\delta$ and the polaritons are a combination of excitons and cavity photons.
	This mixing is due to the strong coupling of excitons and cavity photons. 

\section{Branch-entangled polaritons}\label{Sec:BEPP}
Since we are interested in the generation of entangled polariton pairs, we consider a situation where a pump laser stimulates scattering processes of polaritons.
	It was shown theoretically in Ref.~\cite{PDSSS09} and experimentally in Ref.~\cite{SDSSL05}, using ``which-way'' experiments, that pumping the lower polariton branch can lead to the generation of polarization-entangled polariton pairs.
	Here we are interested in a different situation where the pump laser drives coherently the upper branch at a given wave vector $\mathbf{k}_p$, as illustrated by Fig.~\ref{sketch}(b) for $\mathbf{k}_p=0.05 k_0 \mathbf{e}_x$, $E_C(0) = 1.5\,$eV, $\Omega_R = 2\,$meV and $\delta = 0$.
	The pumped polaritons (solid black circles) scatter into states belonging to different branches, $j_1 \ne j_2$ (open black circles).
	In this setting, frequency or branch entanglement arises since both paths (indicated by the green lines) are indistinguishable, i.\,e., they are simultaneously phase-matched.
	For strong pumping we approximately replace the annihilation operator $p_{2 \mathbf{k}_p}$ by its mean field value $\langle p_{2 \mathbf{k}_p} \rangle$ and use $P_{2 \mathbf{k}_p}^2 = \langle p_{2 \mathbf{k}_p} \rangle^2 R_X^2 / A$.
	This yields a parametric Hamiltonian
	\begin{multline}\label{Hpar}
		H_{PP}^{\text{par}} = \frac{1}{2} \sum_\mathbf{q} P_{2 \mathbf{k}_p}^2 \Big( V_{\mathbf{k}_p,\mathbf{k}_p,\mathbf{q}}^{1222} p_{1 \mathbf{k}_p+\mathbf{q}}^\dagger p_{2 \mathbf{k}_p-\mathbf{q}}^\dagger \\
		+ V_{\mathbf{k}_p,\mathbf{k}_p,\mathbf{q}}^{2122} p_{2 \mathbf{k}_p+\mathbf{q}}^\dagger p_{1 \mathbf{k}_p-\mathbf{q}}^\dagger \Big) + \text{H.c.}
	\end{multline}
	that approximates the polariton-polariton interaction Hamiltonian~\eqref{HPP} for the scattering process of Fig.~\ref{sketch}(b).
	Note that each pair of polariton creation operators has a different effective potential, such that we cannot factor out the effective potential as done in Ref.~\cite{Ciu04}.
	Additionally, there is a mean-field shift of the branch-dependent energy:
	\begin{equation}
		\widetilde{E}_j(\mathbf{k}) = E_j(\mathbf{k}) + \Lambda_{\mathbf{k},\mathbf{k}_p}^{j2} |P_{2 \mathbf{k}_p}|^2\,,
	\end{equation}
	where
	\begin{equation}
		\Lambda_{\mathbf{q},\mathbf{k}_p}^{j2} = \frac{1}{2} \big( V_{\mathbf{q},\mathbf{k}_p,0}^{j2j2} + V_{\mathbf{k}_p,\mathbf{q},0}^{2j2j} + V_{\mathbf{q},\mathbf{k}_p,\mathbf{k}_p-\mathbf{q}}^{2jj2} + V_{\mathbf{k}_p,\mathbf{q},\mathbf{q}-\mathbf{k}_p}^{j22j} \big) \:.
	\end{equation}

	Assuming that the scattering wave vector $\mathbf{q}$ fulfills the phase-matching condition for the considered interbranch polariton pair scattering process,
	\begin{equation}\label{phase}
		E_{2(1)}(\mathbf{k}_p+\mathbf{q}) + E_{1(2)}(\mathbf{k}_p-\mathbf{q}) = 2 E_2(\mathbf{k}_p),
	\end{equation}
	the Hamiltonian $H_{PP}^{\text{par}}$ from Eq.~\eqref{Hpar} applied on the vacuum state $|\text{vac}\rangle$ generates branch-entangled pairs of polaritons in the state
	\begin{equation}\label{psi_pair}
		|\phi\rangle = \Big( \alpha \, p_{1 \mathbf{k}_p+\mathbf{q}}^\dagger p_{2 \mathbf{k}_p-\mathbf{q}}^\dagger + \beta \, p_{2 \mathbf{k}_p+\mathbf{q}}^\dagger p_{1 \mathbf{k}_p-\mathbf{q}}^\dagger \Big) |\text{vac}\rangle \:.
	\end{equation}
	Here we introduced the parameters
	\begin{subequations}\label{alphbet}
		\begin{align}\label{alpha}
			\alpha &= V_{\mathbf{k}_p,\mathbf{k}_p,\mathbf{q}}^{1222} \Big[ \big(V_{\mathbf{k}_p,\mathbf{k}_p,\mathbf{q}}^{1222}\big)^2 + \big(V_{\mathbf{k}_p,\mathbf{k}_p,\mathbf{q}}^{2122}\big)^2 \Big]^{-1/2} \:, \\\label{beta}
			\beta &= V_{\mathbf{k}_p,\mathbf{k}_p,\mathbf{q}}^{2122} \Big[ \big(V_{\mathbf{k}_p,\mathbf{k}_p,\mathbf{q}}^{1222}\big)^2 + \big(V_{\mathbf{k}_p,\mathbf{k}_p,\mathbf{q}}^{2122}\big)^2 \Big]^{-1/2} \:,
		\end{align}
	\end{subequations}
	characterizing the properties of the material.
	In contrast to Ref.~\cite{Ciu04}, the state $|\phi\rangle$ in Eq.~\eqref{psi_pair} is not a Bell state (for $\alpha^2\ne \beta^2$), which is due to the inequality of the effective branch-dependent potentials.
	As usual, $\alpha^2 + \beta^2 = 1$ ensures the normalization of $|\phi\rangle$.

	In that the polariton energy dispersions $E_{1,2}(\mathbf{k})$ depend on $k$ only, the phase-matching condition is fulfilled if
	\begin{equation}\label{dir_q}
		|\mathbf{k}_p+\mathbf{q}|^2 = |\mathbf{k}_p-\mathbf{q}|^2 \:,
	\end{equation}
	being equivalent to $\mathbf{q} \perp \mathbf{k}_p$. The second part of the phase-matching condition in Eq.~\eqref{phase} yields
	\begin{equation}\label{solphase}
		E_C(|\mathbf{k}_p+\mathbf{q}|) + E_X = 2 E_2(\mathbf{k}_p) \:.
	\end{equation}
	The solution of this equation gives the absolute value of the scattering wave vector $\mathbf{q}$:
	\begin{equation}\label{abs_q}
		|\mathbf{q}|^2 = \Big( \frac{2 E_2(\mathbf{k}_p) - E_X}{E_C(0)} \Big)^2 - 1 - |\mathbf{k}_p|^2 \:.
	\end{equation}
	Because the sign of $\mathbf{q}$ remains unspecified, the phase-matching condition is fulfilled for two equivalent interbranch polariton pair scattering processes.
	Entangled polaritons in the state~\eqref{psi_pair} appear due to the indistinguishability of these scattering channels.

	As we have mentioned above, the value of $\beta$ influences the non-local character of $|\phi\rangle$, cf.~Eq.~\eqref{Bell-like}.
	In case $\beta^2 = 1/2$, we have a true Bell state, and $|\phi\rangle$ is separable for $\beta^2 = 0$ or $\beta^2=1$.
	In all other cases, we have an entangled state as a superposition of two product states.
	Such states are referred to as Bell-like states.
	They violate a Bell inequality, but not maximally~\cite{Bel64}.

	\begin{figure}
		\includegraphics[width=0.48\textwidth]{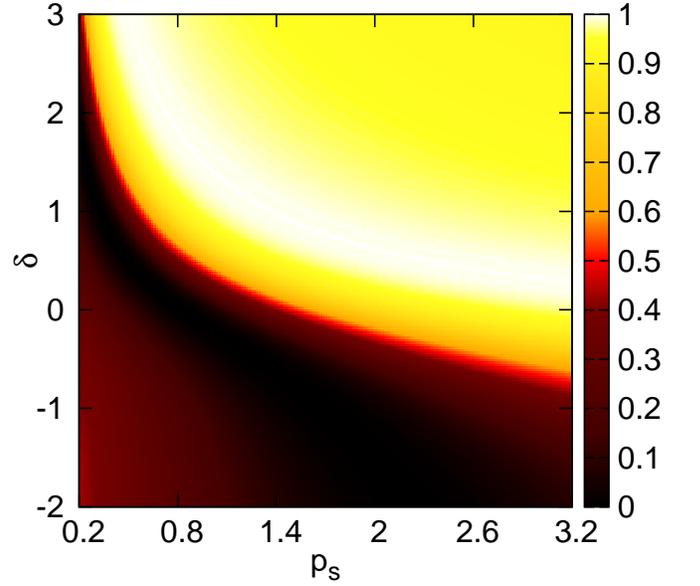}
		\caption{
			(Color online) Magnitude of $\beta^2$ in the $(\delta, p_s)$ plane according to Eq.~\eqref{beta} for $E_C(0)= 1.5\,$eV, $E_b = 10\,$meV and $\mathbf{k}_p = 0.05 k_0 \mathbf{e}_x$.
			The phase-matching scattering wave vector $\mathbf{q}$ follows from Eqs.~\eqref{dir_q} and~\eqref{abs_q}.
		}
		\label{bet}
	\end{figure}
	In Fig.~\ref{bet} we plot the value of $\beta^2$ for the phase-matching scattering wave vector $\mathbf{q}$ following from Eqs.~\eqref{dir_q}~and~\eqref{abs_q} as a function of the normalized detuning $\delta$ and the polariton splitting to binding energy ratio $p_s$.
	From this figure we can deduce that the state described by Eq.~\eqref{psi_pair} is a true Bell state only on a specific line in the $(\delta, p_s)$ plane.
	For values of $\delta$ and $p_s$ apart from this line the state of the polariton pair is an entangled Bell-like state.
	Since $\delta$ and $p_s$ are determined by the material, we are in the position to tune the entanglement properties of the polariton pairs.
	For example, we might consider materials, where the polariton splitting is of the order $2 \Omega_R \sim 4\,$meV, while the exciton binding energy approximately is $E_b \sim 10\,$meV. Since the ratio of the anharmonic 
exciton-photon interaction to the exciton-exciton interaction, $[4 \pi \Omega_R / (21 E_b)]^2$, is of the order of $10^{-2}$, it is a fairly 
good approximation to omit the anharmonic part of the exciton-photon coupling.
The particular choice $p_s = 0$ causes a simpler effective branch-dependent 
potential
	\begin{equation}\label{effpot_ps0}
		\frac{V_{\mathbf{k},\mathbf{k'},\mathbf{q}}^{j_1j_2j_3j_4}}{E_b} \approx 12 M_{1j_1 \mathbf{k}+\mathbf{q}} M_{1j_2 \mathbf{k'}-\mathbf{q}} M_{1j_3 \mathbf{k}} M_{1j_4 \mathbf{k'}} \:,
	\end{equation}
	$V_{\mathbf{k}_p,\mathbf{k}_p,\mathbf{q}}^{1222} = V_{\mathbf{k}_p,\mathbf{k}_p,\mathbf{q}}^{2122}$, and $\alpha^2 = \beta^2 = 1/2$.
	Obviously, such microcavities create polariton pairs in a Bell state configuration.

	Another important effect results when we apply several pumps with different pump wave vectors $\mathbf{k}_{p1}, \mathbf{k}_{p2}, \dots$ to the microcavity.
	Motivated by experiments is a pump-pulse train, where all $\mathbf{k}_{pn}$, $n = 1, 2, \dots$ are aligned in the same direction but have different amplitudes.
	Then we have a phase-matching condition for each $\mathbf{k}_{pn}$.
	Accordingly, branch-entangled polariton pairs appear for all phase-matching scattering wave-vectors $\mathbf{q}_n$ following from Eqs.~\eqref{dir_q} and~\eqref{abs_q} by inserting the respective pump vector $\mathbf{k}_{pn}$.
	Then the state of $N$ branch-entangled polariton pairs takes the form
	\begin{multline}\label{Bell-like}
		|\psi\rangle = \prod_{n=1}^N \Big( \alpha_n \, p_{1 \mathbf{k}_{pn}+\mathbf{q}_n}^\dagger p_{2 \mathbf{k}_{pn}-\mathbf{q}_n}^\dagger \\
		+ \beta_n \, p_{2 \mathbf{k}_{pn}+\mathbf{q}_n}^\dagger p_{1 \mathbf{k}_{pn}-\mathbf{q}_n}^\dagger \Big) |\text{vac}\rangle \:.
	\end{multline}
	The normalization $\prod (\alpha_n^2 + \beta_n^2) = 1$ of this state follows from the property $\alpha_n^2 + \beta_n^2 = 1$ for each $n$.

\section{Entanglement of emitted light}\label{Sec:FreqEntPhot}
\subsection{Frequency-entangled photons}
	In the following, we consider the emission of entangled light from the microcavity.
	As shown in Ref.~\cite{Ciu04}, 
the coupling of the intra-cavity polaritons to an external field 
can be described by the quasimode Hamiltonian
	\begin{equation}
		H_{FP}^{\text{ext}} = \sum_{j,\mathbf{k}} \int {\mathrm d}\omega \, g(\omega) |M_{j2\mathbf{k}}|^2 a_{\omega,\mathbf{k}}^\dagger p_{j\mathbf{k}}^{} + \text{H.c.}
	\end{equation}
	with a frequency-dependent coupling $g(\omega)$.
	The creation operator $a_{\omega,\mathbf{k}}^\dagger$ describes an emitted photon with frequency $\omega$ and in-plane wave vector $\mathbf{k}$.
	The coupling of each branch to the external field is proportional to the photonic fraction $|M_{j2\mathbf{k}}|^2$.
	If the $|M_{j2\mathbf{k}}|^2$ are of comparable magnitude, the branch entanglement of the polaritons transfers to a frequency entanglement of photon pairs in the state
	\begin{equation}\label{psi_alpha}
		|\psi\rangle = \prod_{n=1}^N\left( \alpha_n \, a_{n,-}^\dagger \, a_{\overline{n},+}^\dagger + \beta_n a_{\overline{n},-}^\dagger \, a_{n,+}^\dagger \right) |\text{vac}\rangle,
	\end{equation}
	where the multi-indices are defined as
	\begin{equation}\begin{split}
		(n,\pm) &= \big( E_1(\mathbf{k}_{pn}+\mathbf{q}_n), \mathbf{k}_{pn}\pm\mathbf{q}_n \big) \:, \\
		(\overline{n}, \pm) &= \big( E_2(\mathbf{k}_{pn}+\mathbf{q}_n), \mathbf{k}_{pn}\pm\mathbf{q}_n \big) \:.
	\end{split}\end{equation}
	Obviously, $E_{1(2)}(\mathbf{k}_{pn}-\mathbf{q}_n) = E_{1(2)}(\mathbf{k}_{pn}+\mathbf{q}_n)$, for scattering wave vectors $\mathbf{q}_n$, fulfilling the phase-matching condition~\eqref{phase}.

	We now identify the entanglement of the multiple photon pairs as strong entanglement by changing the point of view according to Fig.~\ref{sketch}(c).
	For this purpose, we decompose the compound Hilbert space $\mathcal H$ of the emitted photons into two parties, $\mathcal H=\mathcal H_-\otimes\mathcal H_+$, where the subspaces $\mathcal H_\pm$ contain all photons emitted with an in-plane wave vector $\mathbf k_{pn}\pm\mathbf q_n$, respectively.
	In Fig.~\ref{sketch}(c), this yields two spatial subspaces for all photons emitted to the left-hand side or to the right-hand side.
	We choose this particular decomposition in order to quantify entanglement in possible which-way experiments.
	It is important to stress that other possible decompositions could be treated similarily, but they may give other outcomes~\cite{HHHH09,ZLL04}. 
  Our particular decomposition is motivated from experimental accessibility.
	Consider, for example, the above mentioned pump pulse train, where all $\mathbf{k}_{pn}$ are aligned in the same direction $\mathbf{e}_p$, but have different amplitudes.
	Then, according to the solution of the phase-matching condition~\eqref{phase} in Sec.~\ref{Sec:BEPP}, all scattering wave vectors $\mathbf{q}_n$ are perpendicular to the symmetry axis $\mathbf{e}_p$, i.\,e. photons with wave vectors $\mathbf{k}_{pn} \pm \mathbf{q}_n$ are spatially separated. 

	Let us describe this decomposition mathematically.
	We may introduce the states $|0\rangle_n$ for a photon with an energy $E_1(\mathbf{k}_{pn}+\mathbf{q}_n)$, and $|1\rangle_n$ for a photon energy $E_2(\mathbf{k}_{pn}+\mathbf{q}_n)$.
	With these definitions we get $a_{n,-}^\dagger a_{\overline{n},+}^\dagger |\text{vac}\rangle = |0\rangle_n\otimes| 1\rangle_n$ and $a_{\overline{n},-}^\dagger \, a_{n,+}^\dagger|\text{vac}\rangle = |1\rangle_n\otimes| 0\rangle_n$.
	Thus, the state $|\psi\rangle$ in Eq.~\eqref{psi_alpha} reads
	\begin{equation}
		|\psi\rangle = \prod_{n=1}^N \Big( \alpha_n |0\rangle_n\otimes| 1\rangle_n + \beta_n |1\rangle_n\otimes| 0\rangle_n \Big).
	\end{equation}

	The expansion of this product yields a sum of $2^N$ product states
	\begin{multline}
		\left(\prod_{n=1}^N|i_n\rangle_n\right)\otimes\left(\prod_{n=1}^N|1-i_n\rangle_n\right)\\
		=|i_1,\dots,i_N\rangle\otimes|1-i_1,\dots,1-i_N\rangle
	\end{multline}
	with $i_n\in\{0,1\}$.
	The sequence $(i_n)_{n=1}^N$ can be understood as a binary representation of an integer $m$ between $0$ and $2^N-1$, whereas the corresponding sequence $(1-i_n)_{n=1}^N$ gives the complement integer $\overline m=(2^N-1)-m$.
	As a result we obtain
	\begin{equation}\label{psi}
		|\psi\rangle=\sum_{m=0}^{2^N-1}\gamma_m|m,\overline m\rangle
	\end{equation}
	with coefficients
	\begin{equation}\label{Gamma}
		\gamma_m=\prod_{n=1}^N\big[(1-i_n)\alpha_n+i_n\beta_n\big].
	\end{equation}
	The expression $(1-i_n)\alpha_n+i_n\beta_n$ equals $\alpha_n$ for $i_n=0$ and $\beta_n$ for $i_n=1$.
	The normalization condition reads 
\begin{equation}\label{nc}
\sum_{m=0}^{2^N-1} \gamma_m^2 = \prod_{n=1}^N (\alpha_n^2 + \beta_n^2)=1\,.
\end{equation}
	Note that in the form of Eq.~\eqref{psi} $|\psi\rangle$ is no longer a multipartite product state, but a strongly entangled bipartite state.

\subsection{Identification of strongly entangled states}\label{Sec:IdentEnt}
	To identify bipartite entanglement we use entanglement witnesses~\cite{HHH96,SV09a}, or, more specifically, SN (Schmidt number) witnesses.
	For pure states the SN arises from the Schmidt decomposition of the state~\cite{NC10}.
	For example, if we consider the pure state $|\psi\rangle$, cf.~Eq.~\eqref{psi}, the SN is the number of nonzero coefficients $\gamma_m$.
	Thus, the SN quantifies the entanglement based on the quantum superposition of the product states $|m,\overline m\rangle$. SN witnesses can 
also be employed for mixed quantum states.

	The construction of SN witnesses is a challenging task.
	Recently we have shown that one can use general Hermitian operators to identify the amount of entanglement~\cite{SV11a}.
	A (in general mixed) quantum state has a SN greater than $r$ if and only if there exists a Hermitian operator $L$ with
	\begin{equation}\label{Eq:SNtest}
		\langle L\rangle=\mathop{\text{Tr}} \rho L > f_r(L) \:,
	\end{equation}
	where
	\begin{equation}\begin{split}
		f_r(L) &= \sup \big\{ \langle\psi_r|L|\psi_r\rangle : |\psi_r\rangle~\text{SN}~r~\text{state} \big\} \:.
	\end{split}\end{equation}
	A SN witness can be constructed from $(f_r(L)\mathbb I-L)$.
	Obviously, the case $r=1$ is equivalent to an entanglement test~\cite{SV09a}.
	A possible way to identify the value of the function $f_r(L)$ is based on a generalized eigenvalue equation---the so-called SN eigenvalue equation---which takes the form
	\begin{align}
		L|\psi_r\rangle=g|\psi_r\rangle+|\chi\rangle
	\end{align}
	with $|\chi\rangle$ being a bi-orthogonal perturbation, cf.~\cite{SV11a}.
	The value $g$ is the SN eigenvalue and the vector $|\psi_r\rangle$ is the SN eigenvector.
	The largest SN eigenvalue is the value of the function $f_{r}(L)$ for the SN test in Eq.~\eqref{Eq:SNtest}.
	The case $r=1$ delivers the separability eigenvalue equations~\cite{SV09a}, and we have shown that they also apply to the identification of entanglement via negative quasiprobabilities~\cite{SV09b,SV12}.

	Now, let us measure the entanglement with respect to the chosen decomposition of the Hilbert space $\mathcal{H}$.
	To determine the SN of the state, we consider the projection $L=|\psi\rangle\langle\psi|$ and obtain $\langle L\rangle=\langle\psi|L|\psi\rangle=1$.
	For the function $f_{r}(L)$ we get
	\begin{equation}
		f_r(L)=\max\{\gamma_{m_1}^2+\dots+\gamma_{m_r}^2: m_i\neq m_j \text{ for } i\neq j\} \:,
	\end{equation}
	which is the sum of the $r$ largest squared Schmidt coefficients~\cite{SV11a}.
	Due to the normalization of the state, $\sum_{m=0}^{2^N-1}\gamma_m^2=1$, the value of $f_r(L)$ is smaller than 1, if there exist more than $r$ values $\beta_m\neq0$.
	In conclusion, the considered pure state $|\psi\rangle$ has a SN of $2^N$, in the general case that all $\alpha_n,\beta_n\neq0$ for $n=1,\dots,N$.

	We conclude that the emitted light, which directly corresponds to the cavity-internal quantum state, is strongly entangled.
	In order to generate such a state, the quantum superposition of local states $|m,\overline m\rangle$, is required at least $r=2^N$ times.
	These strongly entangled outputs verify the internal quantum correlation between the branch-entangled polaritons inside the cavity structure.
	However, in a more realistic scenario, we have imperfections causing a loss of quantum entanglement.
	For example, the initially strongly entangled state $|\psi\rangle\langle\psi|$ could undergo a dephasing.
	In the limiting case of full dephasing, the state $\rho_{\rm deph}$ becomes
	\begin{equation}
		\rho_{\rm deph}=\sum_{m=0}^{2^N-1}\gamma_m^2 |m,\overline m\rangle\langle m,\overline m|,
	\end{equation}
	and contains no interferences of the form $|m,\overline m\rangle\langle l,\overline l|$ for $l\neq m$.
	In this scenario, the SN equals the minimum value one for the separable state $\rho_{\rm deph}$.
	This means that this state is useless for any protocol based on entanglement.
	In the following, we will study the amount of entanglement in the intermediate region between no and full dephasing.

\section{Dephasing}\label{Sec:LinDephase}
	In quantum optics the role of losses is crucial and has to be considered carefully.
	On the one hand there are internal losses leading to branch-, wave-vector-, and excitation-density-dependent broadenings for the polariton modes.
	Examples are scattering with acoustic phonons~\cite{PDSSRG08a,PDSSS09}, mixing with states of the exciton continuum~\cite{CK03}, Coulomb induced parametric scattering~\cite{PDSSRG08a}, or losses through the cavity mirrors.
	On the other hand, there are external losses diminishing the initially available amount of entanglement.
	Once entangled radiation is emitted out of the cavity a major source for the loss of entanglement is dephasing~\cite{SV11a,SV12}.
	We here aim to quantify this lossy channel, i.\,e., we neglect all internal losses and assume that the microcavity emits strongly entangled photons that shall be detected at a certain fixed distance. 

\subsection{Propagation through different linear media}
        In the bipartite setting under study, the two parts of the entangled radiation field would in general propagate through different media, cf.~Fig.~\ref{sketch}(c). 
        In the case of pumping by a pulse train, already some small differences in the dispersive properties of the two media would lead to significant relative phase shifts and hence to an overall dephasing effect diminishing the entanglement between the output channels of the two transmission lines.

	Let us assume two media with linear dispersions given by $\omega_\pm(k)$, where the index $\pm$ indicates the propagation in $\mathcal H_\pm$, respectively.
	The Hamiltonian reads $H_{\rm deph} = H_- + H_+$, where
	\begin{equation}
		H_\pm = \sum_{n=1}^N \Big[ \omega_\pm(E_{1n}) a_{n,\pm}^\dagger a_{n,\pm}^{} + \omega_\pm(E_{2n}) a_{\overline{n},\pm}^\dagger \, a_{\overline{n},\pm}^{} \Big]
	\end{equation}
	with the energies $E_{1n(2n)} = E_{1(2)}(\mathbf{k}_{pn}\pm\mathbf{q}_n)$.
	Recall that energies for wave vectors $\mathbf{k}_{pn} \pm \mathbf{q}_n$ are identical for phase-matching scattering wave vectors $\mathbf q_n$.
	These Hamiltonians are diagonal in the photon number basis, such that
	\begin{equation}
		H_\pm |m\rangle = E_{m,\pm} |m\rangle
	\end{equation}
	with modified eigenvalues in the binary representation $(i_n)_{n=1}^N$ of the integer $m \in [0, 2^N-1]$:
	\begin{equation}
		E_{m,\pm} = \sum_{n=1}^N \Big[ (1 - {i_n}) \omega_\pm(E_{1n}) + {i_n} \omega_\pm(E_{2n}) \Big] \:.
	\end{equation}
	It is obvious that the vacuum can be expressed in the same way using the dispersion relation $\omega_{\rm vac}(k) = k$.

	The time evolution of the initially emitted radiation is
	\begin{equation}\label{psi_t}\begin{split}
		|\psi(t)\rangle &= \rme^{-\rmi(H_- + H_+)t} |\psi\rangle \\
		&= \sum_{m=0}^{2^N-1} \rme^{-\rmi(E_{m,-} + E_{\overline{m},+})t} \gamma_m |m, \overline{m}\rangle \:.
	\end{split}\end{equation}
	The spatial distances from the cavity to detectors in the left and right subspaces are assumed to be equal.
	However, the optical path lengths differ in both parties and depend on the frequency components of the propagating fields.
	Effectively, the arrival times at the detectors differ for the different field components created by the microcavity system.
	This leads to the exponential factor in Eq.~\eqref{psi_t} which takes into account the phase shift between the two parties of photons.

	To obtain the photon state measured by the detectors, we have to average over the different arrival times to account for the different optical path lengths.
        In practice the resulting statistics depends on the details of the dispersive properties of both media representing the two propagation channels.
	Such a treatment must be based on an experimental analysis of the used channels, which is beyond the scope of the present paper.
	To demonstrate the basic principles, we simply suppose an equally distributed difference of the arrival times in the two channels.
	This yields
	\begin{equation}\begin{split}\label{MixedDensity}
		\overline{\rho}(t_1, t_2) =& \frac{1}{t_1 - t_2} \int_{t_1}^{t_2} \rmd t\,|\psi(t)\rangle\langle\psi(t)|\\
		=& \sum_{m,l=0}^{2^N-1} \gamma_m \gamma_l \, \rme^{-\rmi x_{ml} (t_1 + t_2) / 2} \\
		&\times \mathop{\mathrm{sinc}} \Big( x_{ml} \frac{t_2 - t_1}{2} \Big) |m, \overline{m}\rangle \langle l, \overline{l}| \:,
	\end{split}\end{equation}
	with
	\begin{equation}
		\int_a^b \rme^{-\rmi x t} \, \rmd t = (b - a) \rme^{-\rmi x (a+b) / 2} \mathop{\text{sinc}} \Big(x \frac{b-a}{2} \Big) \:,
	\end{equation}
	where $x_{ml} = (E_{m,-} + E_{\overline{m},+} - E_{l,-} - E_{\overline{l},+})$, $\mathop{\text{sinc}}(y) = \sin(y) / y$.
	The state $\overline{\rho}(t_1, t_2)$ in Eq.~\eqref{MixedDensity} represents the structure of the density operator of entangled light suffering from dephasing.
	For $\Delta t = t_2 - t_1 \to \infty$ we obtain the separable state $\rho_{\rm deph}$.
	All the correlations generated by the branch-entangled polaritons vanish in this extremal situation.

	Clearly there is no dephasing if the photons in both Hilbert spaces $\mathcal{H}_\pm$ propagate through the same medium, $\omega_\pm(k)=\omega(k)$, as it is perfectly realized in vacuum channels.
	Under such conditions, the sum
	\begin{equation}
		E_{m,\pm} + E_{\overline{m},\mp} = \sum_{n=1}^N \Big[ \omega(E_{1n}) + \omega(E_{2n}) \Big]
	\end{equation}
	is independent of $m$ such that $x_{ml} = 0$ for all $m,l \in [0,2^N-1]$.
	The difference in the optical path lengths only depends on the difference of the dispersion relations between left and right Hilbert space.
	Thus, without loss of generality, we can assume that we have a free-space propagation in $\mathcal H_+$ and a linear medium in $\mathcal H_-$ as anticipated in Fig.~\ref{sketch}~(c).

\subsection{Detection of strong entanglement}
	To quantify strong entanglement in the continuous-variable mixed state $\overline\rho(t_1,t_2)$ for finite $\Delta t$, we need to find a suitable test operator $L$.
	As we have seen in the case of pure states, the test operator should be closely related to the density operator in order to have a large value on the left-hand side of Eq.~\eqref{Eq:SNtest}.
	On the other hand, the value $f_{r}(L)$ should be as small as possible.
	Together this means that we should use a test operator $L$ in the form
	\begin{equation}
		L=\sum_{m,l=0}^{2^N-1}\Gamma_{m,l}|m,\overline m\rangle\langle l,\overline l| \:,
	\end{equation}
	with the positive semi-definite matrix of coefficients
	\begin{equation}
		\mathbf{\Gamma}=(\Gamma_{m,l})_{m,l=0}^{2^N-1} \:.
	\end{equation}
	As shown in Ref.~\cite{SV11a}, in such a case we can obtain the function $f_r(L)$ just by determining the largest eigenvalue of all $r\times r$ principal submatrices of $\mathbf \Gamma$.

	To construct a suitable test operator we consider the given state $\overline \rho(t_1,t_2)=\sum_{m,l=0}^{2^N-1} \rho_{m,l} |m,\overline m\rangle\langle l,\overline l|$.
	Let $x^{(k)}=(x_{m}^{(k)})_{m=0}^{2^N-1}$ be the $k$th eigenvector of the coefficient matrix $(\rho_{m,l})_{m,l=0}^{2^N-1}$ for a nonzero eigenvalue.
	Then we choose $L$ to be the projector in the subspace spanned by the vectors
	\begin{align}
		|x^{(k)}\rangle=\sum_{m=0}^{2^N-1} x_{m}^{(k)} |m,m\rangle.
	\end{align}
	This means $\Gamma_{m,l}=\sum_{k} x_{m}^{(k)}x_{l}^{(k)\ast}$.
	Analogously to the case of a pure state we obtain
	\begin{align}
		\langle L\rangle=\mathop{\rm Tr}\overline{\rho}(t_1,t_2) L=1.
	\end{align}
	As long as the subspace given by all $|x^{(k)}\rangle$ does not contain a SN $r$ vector $|\psi_r\rangle$, for the projection $L$ holds:
	\begin{align}
		f_{r}(L)=\sup_{|\psi_r\rangle}\langle\psi_r|L|\psi_r\rangle<1.
	\end{align}
	Hence we get a SN greater than $r$ whenever $f_{r}(L)<1=\langle L\rangle$.

	At this point, let us comment on the particular choice of the observable $L$.
	The fact that $L$ is a projection guarantees a high verification rate of the SN test given by~\eqref{Eq:SNtest}.
	The main advantage of using $L$, which depends on $\overline{\rho}(t_1,t_2)$, relates to the appearance of a large mean value $\langle L\rangle$ on the left-hand side of~\eqref{Eq:SNtest}, representing the measurement outcome. 
By contrast, the right-hand side of the SN inequality test, for our choice, takes a comparably small value $f_{r}(L)$ because the projected subspace of $L$, 
by construction, has no SN $r$ state in its range.

	\begin{figure}
		\includegraphics[width=0.48\textwidth]{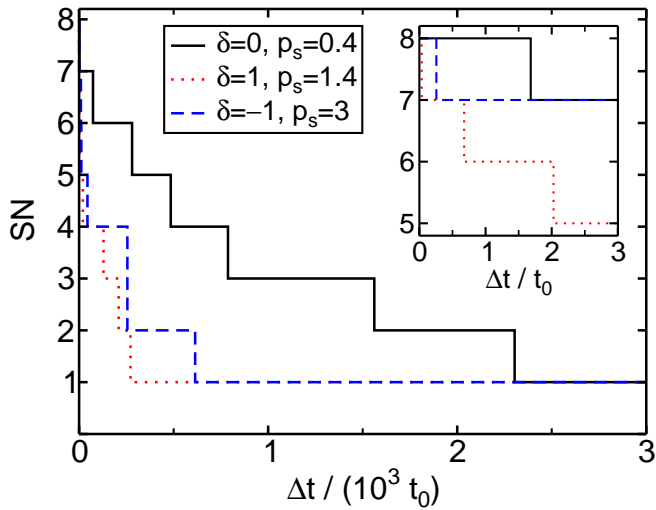}
		\caption{
			Amount of entanglement within the state $\overline{\rho}(t_1,t_2)$ depending on $\Delta t$ and quantized by the SN.
			We apply three different pumps with wave vectors $\mathbf{k}_{pn} = 0.025 n k_0 \mathbf{e}_x$, $n = 1,2,3$ to the cavity.
			The system parameters are $E_C(0) = 1.5\,$eV and $E_b = 10\,$meV.
			The dispersion of the medium is chosen to be $\omega(k) = 0.5 k$.
			Based on our units $\hbar = c = 1$, we choose a typical reference time scale $t_0 = 1\,$eV.
			The inset shows the behavior for weak dephasing.
		}
		\label{deph_SN_N3}
	\end{figure}
	\begin{figure}
		\includegraphics[width=0.48\textwidth]{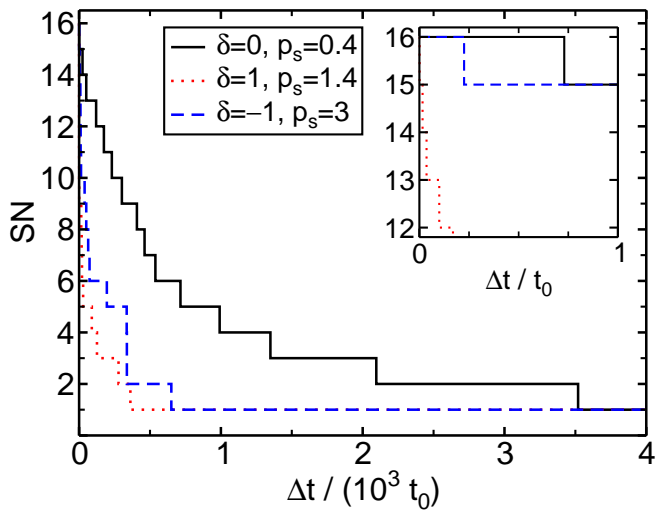}
		\caption{			Amount of entanglement within the state $\overline{\rho}(t_1,t_2)$ for $N = 4$ different pumps.
			The parameters are the same as in Fig.~\ref{deph_SN_N3}, aside from the additional $\mathbf k_{pn}$ for $n=4$.
		}
		\label{deph_SN_N4}
	\end{figure}

	In Figs.~\ref{deph_SN_N3}~and~\ref{deph_SN_N4}, we plot the SN of the state $\overline{\rho}(t_1,t_2)$ depending on $\Delta t$ for different values of the normalized detuning $\delta$ and the ratio of polariton splitting to binding energy $p_s$.
	Figure~\ref{deph_SN_N3} shows the case, where the microcavity is pumped by three beams with different wave vectors aligned in the same direction.
	Hence, the maximal possible SN of the emitted radiation is eight.
	Applying an additional pump, the maximal achievable amount of entanglement increases to 16 (see Fig.~\ref{deph_SN_N4}).

	Both figures indicate that an increasing dephasing due to the increase of $\Delta t$, yields a decreasing SN.
	For $\Delta t = 0$ the SN of the state $\overline{\rho}(t_1,t_2)$ is equal to $2^N$, which is the maximum value.
	The jumps of the value of the guaranteed SN from $r$ to $r-1$ occur for values of $\Delta t$ where the corresponding witness fails to identify a SN larger than $r$.
	For a fixed value of $\Delta t$ the SN strongly depends on the properties $\delta$ and $p_s$ of the planar microcavity.
	A higher number of pump beams---and thus a higher initial SN---may significantly increase the range of $\Delta t$ for which the state $\overline{\rho}(t_1,t_2)$ is still entangled (compare Figs.~\ref{deph_SN_N3}~and~\ref{deph_SN_N4}).

\section{Conclusions}\label{Sec:SaC}
	We have discussed polariton scattering processes within planar semiconductor microcavities with a focus on the possible creation of entangled polariton pairs.
	In extension to previous works, we show that a polychromatic pumping of the upper polariton branch, as motivated by experiments, leads to a simultaneous creation of multiple branch-entangled polariton pairs.
	The coupling of the intra-cavity scattering dynamics to an external field then transfers these kinds of quantum correlations to frequency-entangled photon pairs.
	Since the entanglement properties of these photon pairs are determined by parameters of the device, the measurement of the photon correlations gives valuable information about the internal branch-entanglement within the microcavity.

	The simultaneous creation of photon pairs renders it possible to generate an arbitrary number of copies of entangled qubit states $\rho$, of the form $\rho\otimes\rho\otimes\dots\otimes\rho$.
	Such kinds of states are desired to perform quantum operations based on entanglement, such as quantum teleportation.
	Usually the generation of such states requires that a source of entangled states produces at each time a state $\rho$, which will be stored in a quantum memory to obtain the desired number of copies.
	Here, the number of pump beams or the spectral properties of a pump-pulse train determine the maximal number of simultaneously available entangled qubits.
	By properly choosing the wave vectors of the pump field, one can optimize the Bell type correlations within one or more of those entangled qubits.
	Microcavities pumped with a single pulse of polychromatic light serve as generators of copies of entangled qubit states, making optical quantum memories superfluous.
	Decoherence due to the storage time in a quantum memory cannot occur.

	If desired, the multipartite pair correlations can be mapped to strong bipartite entanglement.
	The quantification of such correlations can be done via the determination of the Schmidt number, which automatically quantifies the multipartite pair correlations and the branch entanglement in the microcavity.
	From our results follows that the Schmidt number of such an unperturbed system is maximal and it can be controlled by the properties of the pump field. A dephasing channel diminishes this resource of entanglement.
	However, we showed that a high amount of entanglement can be guaranteed for a certain range of parameters.
	By using a higher number of pump beams or properly designed pump pulses, one may not only increase the initially available amount of entanglement, but also its resistance against dephasing.

\begin{acknowledgments}
	This work was supported by the Deutsche Forschungsgemeinschaft through SFB 652 by projects B5 and B12.
\end{acknowledgments}

\bibliography{ref}

\end{document}